\documentclass[aps,prb,twocolumn,showpacs,superscriptaddress,floatfix]{revtex4-1}

\usepackage{amsfonts,amsmath,amssymb}
\usepackage{graphicx}

\newcommand{\ket}[1]{|#1\rangle}
\newcommand{\Exp}[1]{\mathrm{e}^{#1}}

\begin{document}

\title{Josephson-Majorana cycle in topological single-electron hybrid  transistors}

\author{Nicolas Didier}
\affiliation{NEST, Scuola Normale Superiore and Istituto Nanoscienze-CNR, I-56126 Pisa, Italy}
\author{Marco Gibertini}
\affiliation{NEST, Scuola Normale Superiore and Istituto Nanoscienze-CNR, I-56126 Pisa, Italy}
\author{Ali G. Moghaddam}
\affiliation{Theoretische Physik, Universit\"at Duisburg-Essen and CENIDE, 47048 Duisburg, Germany}
\affiliation{Department of Physics, Institute for Advanced Studies in Basic Sciences (IASBS), Zanjan 45137-66731, Iran}
\author{J\"urgen K\"onig}
\affiliation{Theoretische Physik, Universit\"at Duisburg-Essen and CENIDE, 47048 Duisburg, Germany}
\author{Rosario Fazio}
\affiliation{NEST, Scuola Normale Superiore and Istituto Nanoscienze-CNR, I-56126 Pisa, Italy}

\begin{abstract}
Charge transport through a small topological superconducting island in contact with a normal and a superconducting
electrode occurs through a cycle that involves coherent oscillations of Cooper pairs and tunneling in/out the normal electrode
through a Majorana bound state, {\it the Josephson-Majorana cycle}.  We illustrate this mechanism by studying the current-voltage
characteristics of a superconductor -- topological superconductor -- normal metal single-electron transistor. At low bias
and temperature the Josephson-Majorana cycle is the dominant mechanism for transport.
We discuss a three-terminal configuration where the non-local character of the Majorana bound states is emergent.
\end{abstract}


\maketitle

\section{Introduction}
\label{introduction}

Majorana bound states are zero-energy states that occur at the boundary or in the vortex core of topological
superconductors.~\cite{qi11,beenakker11,alicea12}  Besides the genuine interest in understanding their properties, they
play a fundamental role in the realization of a  topological quantum computer.~\cite{kitaev01,nayak08} For these reasons
an intense research has started to find physical systems that support Majorana excitations.  A quantum wire with spin-orbit
coupling in close proximity to a superconductor and in the presence of an external magnetic field has been considered
among the most promising proposals.~\cite{lutchyn10,oreg10} The recent experiments~\cite{leo,heiblum,deng,rokhinson}  on this 
system  provide the first evidences of the existence of Majorana bound states in the condensed matter world. 

The presence of Majorana bound states in topological superconductors has a number of distinct signatures.
It is responsible for a  fractional Josephson effect,~\cite{kitaev01,lutchyn10,fu09,jiang11} it leads to a
resonant Andreev reflection~\cite{bolech,nilsson,law,wimmer,flensberg} and to anomalies in interference 
experiments,~\cite{bonderson,stern,strubi} or it can be detected by measuring the local density of states.~\cite{potter,stanescu,sau,black,gibertini}

In the presence of quantum fluctuations, affecting the phase coherence of the topological superconductor,
new effects appear due to a parity constraint linking the dynamics of the superconducting phase and of the
Majorana fermions.  A paradigmatic situation to explore these phenomena, in the setup of Refs.~\onlinecite{kitaev01,
lutchyn10,oreg10}, is when the superconducting island in contact with the wire is small enough so that
charging effects~\cite{grabert-devoret} come into play.~\cite{footnote}
Fu~\cite{fu10} first pointed out that the parity constraint  leads to non-local effects in electron transport. 
The role of charging effects on the fractional Josephson
effect and on the Coulomb blockade through a topological superconducting  transistor was studied in
Refs.~\onlinecite{vanheck11} and ~\onlinecite{zazunov11}, respectively.

\begin{figure}
\includegraphics[width=\columnwidth,clip=true,trim=0 4.5mm 0 7.5mm]{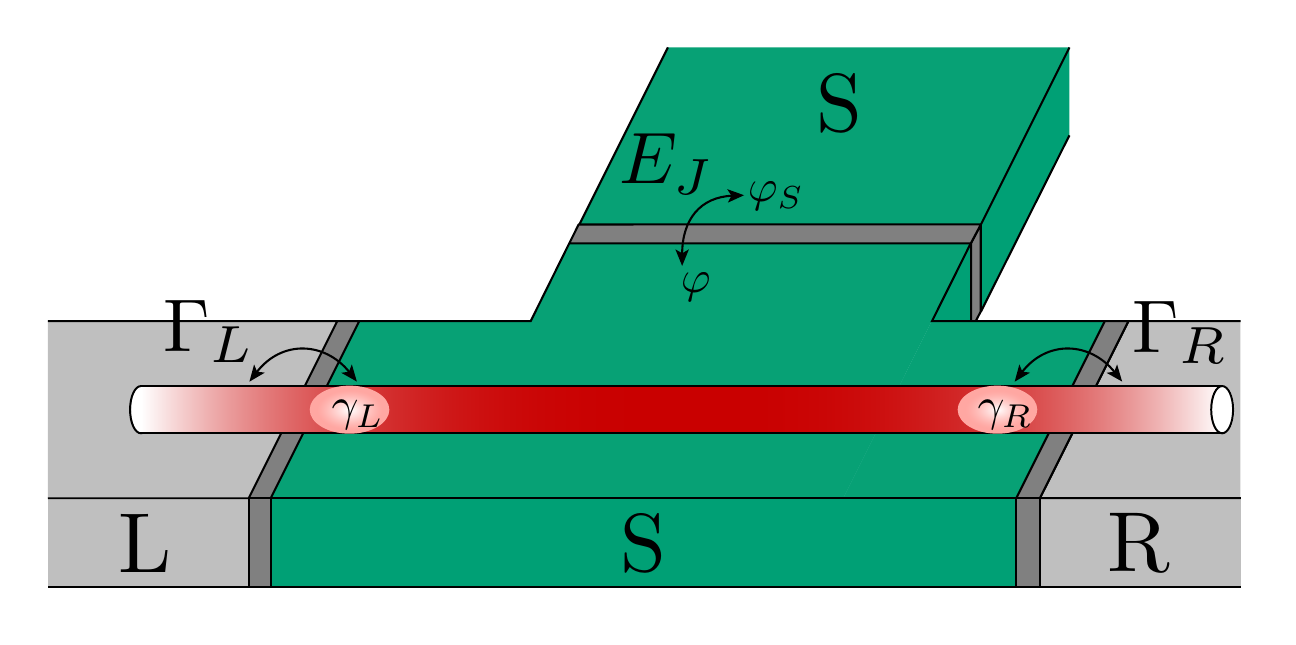}
\caption{The system consists of a superconducting island  connected to two normal metals and to a superconductor. A quantum wire with 
strong spin-orbit interaction is deposited on top of the superconducting island. The whole system is immersed in an external magnetic 
field. By choosing properly the parameters the proximized wire enters a topological phase with two Majorana bound states localized 
at the two ends of the wire.  Tunneling into the normal leads happens through the Majorana states while the coupling of the central island 
to the superconducting lead is due to the Josephson effect.}
\label{setup}
\end{figure}

In this work we introduce and analyze what turns out to be the dominant  charge transport mechanism occurring at low voltages
in hybrid topological single electron transistors: The Josephson-Majorana cycle. It takes place when a topological
superconducting island is coupled to  superconducting and to  normal leads (see Fig.~\ref{setup}).
Charges can flow through the island due to the combined effect of the coherent oscillations of Cooper pairs in the
island and the tunneling between the Majorana state and the normal leads. Although the process bears some similarities
to the Josephson-quasiparticle mechanism present in Cooper pair transistors~\cite{averin89,vandenbrink91,choi01} there
are also important differences.  In the concluding part of the paper we will address this issue and show that some features
of the Josephson-Majorana cycle are true consequences of the non-local character of the Majorana bound states.

The paper is organized as follows. In the next section we describe the setup,  illustrated in Fig.~\ref{setup}, and introduce the underlying Hamiltonian 
describing its dynamics. In Sec.~\ref{cyclesec} we compute the current at low bias through a master-equation approach governing the behavior of 
the reduced density matrix of the central island. In the Coulomb blockade regime the current is dominated by cotunneling. These processes will be 
discussed in Sec.~\ref{cotunneling}. Non-locality of the zero-energy modes will show up only at this higher order in tunneling.
We conclude in Sec.~\ref{conclusions}.

\section{The topological  single-electron hybrid  transistor} 
\label{model}

The system we consider is illustrated in Fig.~\ref{setup}. It consists of a topological superconducting island (a nanowire in close proximity
to a superconducting island) tunnel-coupled  to normal and  superconducting electrodes in a three-terminal configuration.
The island hosts two Majorana bound states at the ends of the wire associated with the Majorana operators $\gamma_{i}$ ($i=L,R$),
$\{\gamma_i,\gamma_j\}=\delta_{i,j}$ and $\gamma_i=\gamma_i^{\dag}$.

At low energies
(much smaller than the superconducting gap), the island couples to the superconducting lead
only through Josephson tunneling. The coupling to the normal leads occurs only via the Majorana bound
states. Throughout the paper we set the two normal electrodes at the same voltage, the only net current will be from the
superconductor to the normal electrodes. The three-terminal setup will, however, be crucial to discuss the non-local character of the
Josephson-Majorana cycle.
The mesoscopic scale of the central island considered here requires to take into account the charging energy of the island.
The Hamiltonian of the system reads:
\begin{equation}
    {\cal H} = {\cal H}_M +  {\cal H}_{Ch} + {\cal H}_{J} + {\cal H}_{leads} + {\cal H}_T~,
\label{hamiltonian}
\end{equation}
where the five terms describe coupling of Majorana states, island charging energy, Josephson coupling to the superconducting electrode,
metallic normal electrode, and tunneling to the normal lead, respectively.
The coupling between the Majorana fermions is given by ${\cal H}_M = i\lambda \gamma_L\gamma_R$ where $\lambda$ is related to the  overlap
between them, which is exponentially small for distances much larger than the superconducting coherence length.
Two Majorana states can form a zero-energy fermionic level described by an annihilation operator $d=(\gamma_L+i\gamma_R)/\sqrt{2}$
which can be either occupied or empty. Therefore, not only the number of excess Cooper pairs $N$ but also the occupation of the $d$-level enters
the charging energy of the island,
\begin{equation}
    {\cal H}_{Ch}= E_C(2N+n_d-n_g)^2~,
\label{charging}
\end{equation}
where $n_d=d^\dag d$ counts the occupation of the $d$-level, $n_g$ is the gate charge  which can be varied continuously by changing
the gate voltage, and $E_C = e^2/(2C)$ is the charging energy expressed in terms of the total capacitance $C$ of the island.
For later convenience, we will label the
eigenstates of the charging energy with the notation $|N,n_d \rangle$.
To describe the Josephson coupling to the superconducting electrode we use the effective Hamiltonian
\begin{equation}
{\cal H}_{J}= -E_J\cos\left(\varphi_S-\varphi\right)~,
\end{equation}
where $E_J$ is the Josephson coupling energy and $\varphi_S$ and $\varphi$ are the phase of the superconducting electrode and the
island condensate respectively.
Here, any possible coupling to the Majorana state can be safely ignored.~\cite{zazunov12}
The normal lead Hamiltonian is given by the noninteracting model 
\begin{equation}
{\cal H}_{leads} = \sum_{i=L,R} \sum_k\epsilon_k c_{k,i}^\dag c_{k,i}
\end{equation}
with creation (annihilation) operator
$c^{\dagger}_{k,i}\;(c_{k,i})  $ corresponding to a spinless fermion in the free particle state $k$ with energy $\epsilon_k$ inside the
left/right (L/R) normal lead.
Finally, the tunneling through the Majorana state takes the form,~\cite{zazunov11}
\begin{equation}
    {\cal H}_T =   \sum_{i=L,R} \sum_k t_{k,i} c_{k,i}^\dag(d+\delta_i e^{-i\varphi}d^\dag)+ {\rm H.c.}~,
\label{tunneling}
\end{equation}
where $t_{k,i}$ is the hopping amplitude to the $k$-state in the $i$-th lead and $\delta_{L/R} =\pm 1$.  The advantage of this formulation introduced by Zazunov
{\it et al.}~\cite{zazunov11}  is the fact that it includes automatically the
constraint on the Hilbert space, linking the occupation of the Majorana bound state to the parity of the superconducting condensate.

A non-trivial dynamics in the problem arises since $\varphi$ and $N$ are canonically conjugated variables, $[N, e^{i\varphi}] = e^{i\varphi}$.
The two terms in the tunneling Hamiltonian correspond to regular and anomalous tunneling. The first one describes the transfer of an
electron from the $d$-level to the normal lead, and the second one the annihilation of a Cooper pair inside the island accompanied by the creation
of two electrons, one in the $d$-level and one in a normal electrode.

\section{Josephson-Majorana cycle and current-voltage characteristics}
\label{cyclesec}

\begin{figure}
\includegraphics[width=0.68\columnwidth]{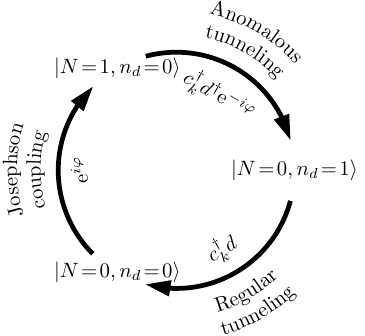}
\caption{Josephson-Majorana cycle for $n_g\sim 1$. The states differing by one
    Cooper pair, $|0,0 \rangle$ and $|1,0 \rangle$, are coupled via the Josephson term. 
    The coupling between $|0,1 \rangle$ and $|0,0 \rangle$ is given by the regular term in the tunneling Hamiltonian,
    the connection between $|0,1 \rangle$ and $|1,0 \rangle$ by the anomalous one.}
\label{cycle}
\end{figure}

We start by considering the current-voltage characteristics at second order in the tunneling amplitudes $t_{k,i}$.
At bias voltage and temperature smaller than the superconducting gap, quasiparticle tunneling is suppressed and coherent
Josephson (Cooper pair) tunneling is necessary for transport through the transistor.
From Eq.~(\ref{charging}) it follows that the resonance condition for coherent Cooper pair oscillations between two charge states
that differ by one Cooper pair is fulfilled at integer values of the gate charge $n_g$.
Around $n_g\sim 1$, e.g., the lowest energy states are $|N,n_d \rangle = |0,0 \rangle$, $|0,1 \rangle$ and  $|1,0 \rangle$.
A Josephson-Majorana cycle involving these three states is illustrated in Fig.~\ref{cycle}.
Starting from $|0,1 \rangle$, a regular tunneling process releases an electron into the (left or right) normal lead.
Then, the island is recharged with an extra Cooper pair provided by the Josephson coupling.
A second (anomalous) tunneling which annihilates a Cooper pair in order to create an electron inside the normal electrode and
another one filling the $d$-level, completes the cycle.
(A cycle with the reverse direction can be obtained by the conjugates of each tunneling process.)
Since the energy of the state $|0,1 \rangle$ is lower than the ones connected by Josephson tunneling, there will be a threshold
voltage for the onset of current. In the following we will support this simple picture sketched above with more detailed calculations.

As already mentioned, the two normal leads are kept at the same voltage.
The average of the total (summed over the two normal leads) current operator, $I=I_L + I_R= ie \,\sum_{i=L,R}\sum_k[c_{k,i}^\dag c_{k,i},H] $, 
can be expressed as 
\begin{equation}
I=-2e\,\mathrm{Im}\, \sum_{i=L,R} \sum_k t_{k,i} c^{\dagger}_{k,i} (d+\delta_i e^{-i\varphi}d^{\dagger}) \;\; .
\end{equation}
 To second order in the tunnel couplings the average current can be conveniently expressed in terms of the reduced density 
 matrix $\rho$ of  the topological superconducting island which is obtained after tracing out the fermionic degrees of freedom of the 
 normal metals. 

The steady-state current is
\begin{multline}
    \langle I \rangle  =  \frac{e}{2} \sum_{i=L,R}  \sum_{l,n,m} \Gamma_i \left[ D^{(i)\ast}_{nl}D^{(i)}_{ml} (F_{ln}+ F_{lm}^*)
    \right. \\ \left.
    -D^{(i)}_{ln}D^{(i)\ast}_{lm}  (\bar{F}_{nl}^*+\bar{F}_{ml} ) \right] \rho_{nm}\,,    \label{exprcurrent}
\end{multline}
where $D^{(i)}_{nm}$ and $\rho_{nm}$ are the matrix elements of the operators $D^{(i)}=d+\delta_i e^{-i\varphi}d^{\dagger}$ and $\rho$ in the basis defined
by the eigenstates $|\psi_n \rangle$ of the Hamiltonian ${\cal H}_{Ch} + {\cal H}_{J}$ with eigenvalues $\mathcal{E}_n$.
The coefficients $F_{nm}$ are defined as
\begin{equation}
 F_{nm}=f(\Delta_{nm}) - \frac{i}{\pi} \, \mathrm{Re}\,\Psi \left( \frac{1}{2}+i \frac{\Delta_{nm}-eV}{2\pi k_BT} \right)\,,
\end{equation}
 with  $\bar{F}=1-F$, $\Delta_{nm}=\mathcal{E}_n-
\mathcal{E}_{m}$, $f(\epsilon)=1/[1+e^{(\epsilon-eV)/k_BT}]$  the Fermi function of the normal electrode,
and $\Psi$  the digamma function.
Finally $\Gamma_i=2\pi|\tau|^2N(\epsilon_F)$ is the tunneling rate with $t_{k,i}\sim \tau$ assumed constant close to the Fermi energy
$\epsilon_F$, and $N(\epsilon_F)$ is the density of states in the normal metal (supposed to be equal for both electrodes). For simplicity 
we will assume that also the interfaces are identical, therefore $\Gamma_L \sim \Gamma_R = \Gamma /2$.

A convenient way to represent and compute the reduced density matrix, in particular when higher-order tunneling processes are taken into
account, is to use a real-time diagrammatic technique that has been developed to describe transport through a metallic single-electron transistor.~\cite{schoeller94,koenig98}
In the absence of the Josephson coupling, $E_J=0$, the eigenstates $|\psi_n \rangle$ are defined by the total island charge 
$|2N+n_d\rangle \equiv |N,n_d \rangle$. Formulated in this basis, the diagrammatic rules for calculating the kernels 
$W_{nm,n'm'}$ are the same as given in Ref.~\onlinecite{koenig98} but with a different rate function, $\alpha^+(\omega) 
\rightarrow (\Gamma/2\pi) f(\omega)$ and $\alpha^-(\omega) \rightarrow (\Gamma/2\pi) [1-f(\omega)]$ and 
extra rules for the overall sign: each crossing of tunneling lines yields a factor $-1$ and, furthermore, the factor $\delta_{L/R}$ needs to be included. 
The transformation of the diagrams into the eigenbasis of 
$\mathcal{H}_{Ch}+\mathcal{H}_{J}$ for finite $E_J$ is straightforward.
The time evolution of the reduced density matrix $\rho(t)$ can be cast in the form 
\begin{equation}\label{rhodot}
\dot{\rho}_{nm}+i\Delta_{nm}\rho_{nm}=\sum_{n',m'}\int_{-\infty}^{t} \mathrm{d}t'W_{nm,n'm'}(t,t'){\rho}_{n'm'}(t')
\, .
\end{equation}
Expanding to second order in the tunneling, this yields for the steady-state limit ($\dot \rho=0$)~\cite{footnote-MEA},
\begin{align}
\nonumber
        &  i\Delta_{nm}{\rho}_{nm} = \frac{\Gamma}{4} \sum_{i,n',m'} \left[
        D^{(i)\ast}_{n'n} D^{(i)}_{m'm} (F_{mn'}+ F_{nm'}^{\ast} ) \right. \\
\nonumber & \left. \hspace{18.5mm}+D^{(i)}_{nn'}D^{(i)\ast}_{mm'} (\bar{F}_{n'm}^*+\bar{F}_{m'n}  ) \right] {\rho}_{n'm'}\\
\nonumber &  -\frac{\Gamma}{4}  \sum_{i,l,n'}  \left[    D^{(i)}_{nl}D^{(i)\ast}_{n'l} F_{lm}^{\ast}  +
        D^{(i)\ast}_{ln}D^{(i)}_{ln'} \bar{F}_{ml}   \right]  {\rho}_{n'm} \\
               &   -\frac{\Gamma}{4} \sum_{i,l,m'} \left[    D^{(i)}_{m'l}D^{(i)\ast}_{ml} F_{ln}   +
               D^{(i)\ast}_{lm'}D^{(i)}_{lm} \bar{F}_{nl}^{\ast}   \right]    {\rho}_{nm'} \, .
        \label{rate}
\end{align}

Having fixed a number of relevant charge states, Eq.~(\ref{rate}) can be solved, together with the normalization condition $\sum_n \rho_{nn} =1$ 
for the reduced density matrix. This solution can be then used to compute the steady-state value of the current as a function of bias and gate voltages. We remark that the current is antisymmetric under the transformation $(V,n_g)\to(-V,2-n_g)$.
To obtain this antisymmetry, we note that the transformation results in the substitutions $F_{nm}\to \bar{F}^{\ast}_{mn}$ and $D_{nm}\to 
D^{\ast}_{mn}$. From Eqs.~(\ref{exprcurrent}) and (\ref{rate}) we conclude that $\rho$ remains unchanged while the current changes sign.

\begin{figure}
\includegraphics[width=0.9\columnwidth]{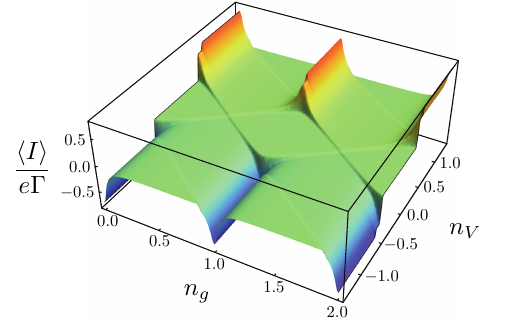}
\caption{Current-voltage characteristics for $k_BT=E_C/100$, $\Gamma=k_BT/10$, $E_J=E_C/5$. In the lowest order in the transmission the 
current is different from zero only in the region $n_g \sim 1$ where two states coupled by the Josephson matrix element are nearly degenerate. 
Outside the region of maximum current the threshold voltage $V^-$ is visible as a tiny line in the Coulomb blockade region. }
\label{iv-gate3d}
\end{figure}

In the range of the gate charges where only the three charge states $|0,0 \rangle$, $|0,1 \rangle$ and $|1,0 \rangle$ are involved, 
a simple analytical solution can be obtained for zero temperature when neglecting the imaginary parts of $F_{nm}$, which are associated 
with energy renormalization induced by the coupling to the bath. In this case the relevant eigenstates and eigenvalues are given by 
\begin{eqnarray}
\ket{\psi_0}&=&\cos\theta\,\Exp{i\varphi_L/2}\,\ket{0}+\sin\theta\,\Exp{-i\varphi_L/2}\,\ket{2} \nonumber\\
\ket{\psi_1}&=&\ket{1} \\
\ket{\psi_2}&=&-\sin\theta\,\Exp{i\varphi_L/2}\,\ket{0}+\cos\theta\,\Exp{-i\varphi_L/2}\,\ket{2} \nonumber
\label{eigenstates}
\end{eqnarray} 
and by
\begin{eqnarray}
\mathcal{E}_0&=&\tfrac{1}{2}(E_0+E_2)-\tfrac{1}{2}\sqrt{(E_2-E_0)^2+E_J^2} \nonumber \\
\mathcal{E}_1&=&E_1 \\
\mathcal{E}_2&=&\tfrac{1}{2}(E_0+E_2)+\tfrac{1}{2}\sqrt{(E_2-E_0)^2+E_J^2}\nonumber 
\label{eigenvalues}
\end{eqnarray} 
respectively. In the previous expression $E_0 = E_C\,n_g^2$, $E_1= E_C(1-n_g)^2$ and $E_2= E_C(2-n_g)^2$,
and the angle $\theta$ is defined by $\tan2\theta= E_J/(E_2-E_0)$.

The current, as a function of bias voltage, increases in a stepwise fashion at two thresholds $V^{\pm}$ defined as 
\begin{equation}
eV^{\pm}=E_C \left[1\pm2\sqrt{(1-n_g)^2+(E_J/4E_C)^2} \right]~.
\label{eqthreshold}
\end{equation}
For positive voltage $V>0$ and $1/2<n_g<3/2$, the steady-state current is \begin{equation}
\langle I \rangle=e\Gamma E_J^2\left\{
\begin{aligned}
&2/[\Gamma^2+ 4\mathcal{E}^2- E_J^2] &V>V^+ \\
&1/[\Gamma^2g_c(n_g) + 2 \mathcal{E}^2] &|V^-|<V< V^+ \\
&0 &V<|V^-|
\end{aligned}\right.
\label{eqIVC}
\end{equation}
where $\mathcal{E} = \sqrt{[4E_C(1-n_g)]^2+E_J^2}$, $g_c(n_g) = (21+13~x-x^2-x^3)/16$, and $x = 4E_C(1-n_g)/\mathcal{E} $. 
We immediately see that the current vanishes when the Josephson coupling goes to zero.
The width of the resonance peak is $\delta n_g=E_J/4E_C$ and the peak value for $V>V^{+}$ becomes maximal 
for $\Gamma=\sqrt{4 \mathcal{E}^2-E_J^2}$. Going beyond the three-state approximation requires a numerical solution of the 
master equation. In Fig.~\ref{iv-gate3d} we show the current as a function of the gate charge $n_g=CV_g/e$ and the bias charge $n_V=CV/e$.
The simple analytical expression given above fully captures all the properties of the current-voltage characteristics. 
The current increases stepwise (smeared by temperature) with the bias voltage and shows resonance peaks at integer values of $n_g$.

\begin{figure}
\includegraphics[width=0.9\columnwidth]{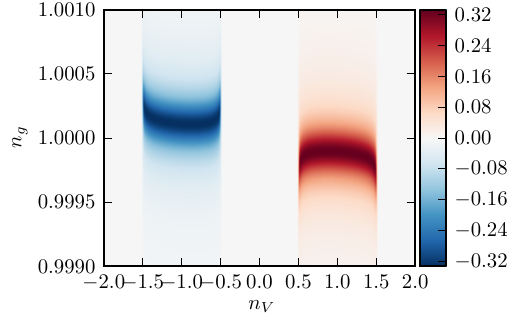}
\caption{Current-voltage characteristics and its dependence on the bias charge $n_V=CV/e$ and the gate charge $n_g=CV_g/e$.
Parameters are: $k_BT=E_C/100$, $\Gamma=k_BT/10$, $E_J=\Gamma/\sqrt{3}$.
The threshold voltage to observe a current is $n_V\simeq\pm0.5$.
The slight displacement of the resonance condition from $n_g=1$ is due to the Lamb shift from the normal leads.}
\label{iv-gate}
\end{figure}

The effect of energy renormalization terms (Lamb shifts) that were neglected in the analytical formulas presented above, is highlighted in 
Fig.~\ref{iv-gate} where the resonance condition is slightly displaced from the value $n_g=1$.
Finally, we comment on the abrupt suppression of the current for large bias voltage $|n_V| > 3/2$ that is visible in Fig.~\ref{iv-gate}.
Beyond the threshold $n_V = 3/2$, the new charge state $|1,1\rangle$ becomes available from $|1,0\rangle$ by tunneling.
The island is then trapped in this new state since the Josephson coupling to the state $|0,1\rangle$ is suppressed because of the large energy difference
of $|1,1\rangle$ and $|0,1\rangle$. As a consequence, transport is blocked.

As already anticipated in the introduction we now discuss the differences between the Josephson-Majorana cycle and the Josephson-quasiparticle 
cycle~\cite{averin89} of Cooper pair transistors. 
An obvious difference, but important for experimental detection, is the voltage scale at which the Josephson-Majorana cycle takes place. Here there 
are no excitations above the gap and therefore the threshold  voltage is set by the charging energy solely.
Second, we note that the Josephson-Majorana cycle shows an $e$-periodicity with respect to $n_g$.\cite{foot}

To give a further characterization of the Majorana states we need to investigate their non-local character. 
In the sequential-tunneling approximation this goal cannot be achieved.
One therefore needs to make one step further and analyze  cotunneling processes where signatures of non-locality first appear.
To this aim it is important to exploit the fork configuration of Fig.~\ref{setup}.

\section{Cotunneling processes}
\label{cotunneling}

It is worth stressing once more that the two metals are kept at the same voltage and transport occurs between the superconducting 
lead and the normal electrodes.
Below the threshold voltage $V^-$, second-order transport considered so far is suppressed by Coulomb blockade.
Nevertheless  fourth-order (two-particle) processes sustain a finite current. Let us consider for illustration the case
$n_g\sim 1$ where the Coulomb threshold is largest. Here the system remains in its ground state $|0,1\rangle$ and
the states $|0,0\rangle$ and $|1,0\rangle$ needed to complete the Josephson-Majorana cycle are only virtually occupied.
Let us consider for the sake of clarity the simple three-charge states case discussed previously. 
To evaluate the current, we determine the cotunneling rates for transitions in which two electrons are transferred from the superconducting 
to the normal leads (or vice versa) while leaving the island in state $|0,1\rangle$.
As intermediate (virtual) states, we take the two linear combinations of $|0,0\rangle$ and $|1,0\rangle$ that form the eigenstates of ${\cal H}_{Ch}+{\cal H}_J$.
All processes that differ only in their intermediate but not initial and final states need to be added coherently.
This includes pairs of processes in which the order of the tunneling events of the two normal lead electrons is interchanged.
Thereby, it is important to distinguish local from nonlocal cotunneling.
For local cotunneling, the amplitudes of those pairs carry opposite sign. 
For nonlocal cotunneling, however, this sign is compensated by the relative sign of $\delta_L=-\delta_R$ in Eq.~(\ref{tunneling}).
As a result, local cotunneling is strongly suppressed, and nonlocal cotunneling dominates.

At $n_g=1$ we get $\theta = \pi/4$ in Eq.~(\ref{eigenstates}), and the two relevant eigenstates are the symmetric and antisymmetric combination of Cooper pair states.
The non-local contribution $I_{LR}$ (Left-Right) to the current is given by 
\begin{equation}
\langle I \rangle_{LR}=e \frac{\Gamma_L \Gamma_R}{\pi} \int_{-eV}^{eV} d\omega \left[  \frac{\mathcal{E}_0} {\mathcal{E}_0^2-\omega^2} -  
\frac{\mathcal{E}_2} {\mathcal{E}_2^2-\omega^2} \right] ^2
\label{nonlocal}
 \end{equation}
On the other side the local (on the same lead) contribution to the current is given by  \begin{equation}
\langle I\rangle_{ii} = e\frac{\Gamma_i^2}{2\pi} \int_{-eV}^{eV} d\omega \left[  \frac{\omega} {\mathcal{E}_0^2-\omega^2} -  
\frac{\omega} {\mathcal{E}_2^2-\omega^2} \right] ^2
\label{local}
\end{equation}
for $i=L,R$. 
 
The local and non-local contributions to the current behave quite differently. 
For zero temperature and to lowest order  in $eV$, we find that 
\begin{equation}
  \langle I \rangle_{LR} =  e \frac{2\Gamma_L \Gamma_R}{\pi} \frac{E_J^2}{(E_C^2-E_J^2/4)^2} eV
\end{equation} 
and, for $i=L,R$,
\begin{equation}
  \langle I \rangle_{ii} =  e \frac{4\Gamma_i^2}{3\pi}  \frac{E_C^2E_J^2}{(E_C^2-E_J^2/4)^4} (e V)^3\, .
\end{equation} 
These results obviously extent to any integer value of $n_g$. At low voltages the local contribution to the current is strongly suppressed. The results
given in Eqs.~(\ref{nonlocal}) and (\ref{local}) are distinct signatures of the nature of Majorana bound states. The dominant contribution of the 
crossed Andreev reflection because of Majorana bound states has been discussed in literature~\cite{nilsson} in the absence of interaction. 
Here we showed that in the Coulomb blockade regime this effect manifests in the cotunneling regime and that it appears as a different voltage 
dependence of the current (linear vs cubic) of the non-local and local contributions respectively.
The difference between local and non-local transport does not appear to lowest order since, there, tunneling of different lead 
electrons occurs incoherently. 

The non-local character of the Majorana states can be detected by the finite linear cotunneling conductance. 
Furthermore, using a counting field at the level of the quantum master equation~\cite{choi01} to derive the statistics of the current leads to a maximally correlated noise, 
\begin{equation}
\langle \delta I_R^2 \rangle = \langle \delta I_L^2 \rangle = \langle \delta I_R \delta I_L \rangle \;, 
\label{maxcor}
\end{equation}
independently of the length of the wire.~\cite{nilsson}

In the absence of Majorana bound states at the ends of the wire there would be a local contribution to the current due to the Andreev processes between 
the normal electrodes and the superconducting island.
These processes would be resonant at $n_g=1$ and, therefore, would contribute to the linear part of the current-voltage 
characteristics.~\cite{hekking} 
In the topological phase, however, the presence of the Majorana bound states changes the situation drastically.  
In this case, as we already remarked, the ground state is $|0,1\rangle$ and the two-electron processes discussed in Ref.~\onlinecite{hekking} are Coulomb blocked.
At low voltage bias, the cotunneling conductance is thus local in the trivial phase and non-local in the non-trivial topological phase. A measurement of the 
cross-correlation noise~\cite{lefloch} would be able to discriminate between the two situations.

\section{Conclusions}
\label{conclusions}

When the superconducting island contacted to the quantum wire is reduced in dimensions, charging effects become important and 
may affect the dynamics of the system in the Majorana sector.  Charging effects in the presence of Majorana bound states can lead 
to a rich phenomenology. In this work we considered a case in which the contemporary presence of tunneling into normal leads 
and the Josephson coupling to a superconducting electrode can provide a new transport mechanism similar to the Josephson-quasiparticle
cycle discussed in superconducting transistors.

In topological hybrid transistors charge transport is dominated at low voltages by a Josephson-Majorana cycle. In this work we analyzed the 
current-voltage characteristics of a topological superconducting island connected to a superconducting and two normal leads. When a 
bias is applied to the leads (both metals are kept at the same voltage) current flows due to this mechanism. It consists in a process in 
which the Rabi oscillations of Cooper pairs are accompanied by the tunneling of electrons from/to Majorana bound
states. To lowest order (sequential tunneling) the Josephson-Majorana cycle leads to a stepwise current-voltage characteristics modulated by the 
gate voltage.  The current is maximum at integer values of $n_g$ when the two charge states differing by one Cooper pair are almost 
degenerate. 

In the sequential-tunneling limit the cycle can reveal the existence of zero-energy states but cannot ascertain their non-local character. 
This fundamental property of Majorana bound states appears in the next order (cotunneling) in the transmission. At this order the three-terminal setup of Fig.~\ref{setup} has two different contributions to the current. There is a  local one in which the two electrons forming the 
Cooper pair tunnel in the same normal lead. In addition, there is a non-local contribution where the Cooper pair is split and tunnels through a 
``virtual'' cycle into the two different leads. This process corresponds to crossed Andreev tunneling. When the central island is in the topological 
phase the local contribution is strongly suppressed (the current is proportional to $V^3$) and only the non-local contribution gives rise to the linear part of the current-voltage characteristics. 

\vspace{0.5cm}
{\it Acknowledgments} --- We acknowledge stimulating discussions with M. Polini and F. Taddei. We are grateful to R. Egger for useful comments.
The work was supported by EU through projects IP-SOLID, QNEMS, GEOMDISS, and NANOCTM and by DFG through Project No. KO 1987/5.


\begin{thebibliography}{77}
\bibitem{foot}   
    If the Majorana level were replaced by a regular  spin-degenerate fermionic level, the current would be $2e$-periodic 
    due to an even-odd asymmetry of the current-carrying states and the corresponding cycles.
\end{thebibliography}
\end{document}